\documentclass[runningheads]{llncs}
\usepackage{graphicx}
\usepackage{comment}
\usepackage{amsmath,amssymb} 
\usepackage{color}
\usepackage{multirow}
\usepackage{diagbox}
\usepackage{caption}
\usepackage{booktabs}
\usepackage{array}
\usepackage{gensymb}
\makeatletter
\newcommand{\thickhline}{%
	\noalign {\ifnum 0=`}\fi \hrule height 1pt
	\futurelet \reserved@a \@xhline
}
\newcolumntype{"}{@{\hskip\tabcolsep\vrule width 1pt\hskip\tabcolsep}}
\makeatother
\begin{document}
	\title{Improving Generalizability in Limited-Angle \\ CT Reconstruction with Sinogram Extrapolation}
	%
	%
	\newcommand{\sign}[1]{\mathrm{sgn}(#1)}
	\author{Ce Wang\inst{1,2} \and
		Haimiao Zhang\inst{3} \and
		Qian Li\inst{1,2} \and
		Kun Shang\inst{4} \and 
		Yuanyuan Lyu\inst{5} \\
		Bin Dong\inst{6} \and
		S. Kevin Zhou\inst{1,7} }
	
	\institute{Key Lab of Intelligent Information Processing of Chinese Academy of Sciences (CAS), Institute of Computing Technology, CAS, Beijing, China \\ \and
		Suzhou Institute of Intelligent Computing Technology, Chinese Academy of Sciences, Suzhou, China  \\ \and
		Institute of Applied Mathematics, Beijing Information Science and Technology University, Beijing, China  \and
		Research Center for Medical AI, Shenzhen Institutes
		of Advanced Technology, Chinese Academy of Sciences, Shenzhen, China  \and
		Z$^{2}$Sky Technologies Inc., Suzhou, China \and
		Beijing International Center for Mathematical Research, Peking University, China  \and 
		Medical Imaging, Robotics, and Analytic Computing Laboratory and Engineering (MIRACLE) School of Biomedical Engineering \& Suzhou Institute for Advanced Research, University of Science and Technology of China, Suzhou 215123, China
		\email{s.kevin.zhou@gmail.com}
	}
	
	\maketitle              
	\begin{abstract}
		
		Computed tomography (CT) reconstruction from X-ray projections acquired within a limited angle range is challenging, especially when the angle range is extremely small. Both analytical and iterative models need more projections for effective modeling. Deep learning methods have gained prevalence due to their excellent reconstruction performances, but such success is mainly limited within the same dataset and does not generalize across datasets with different distributions. Hereby we propose ExtraPolationNetwork for limited-angle CT reconstruction via the introduction of a sinogram extrapolation module, which is theoretically justified. The module complements extra sinogram information and boots model generalizability. Extensive experimental results show that our reconstruction model achieves state-of-the-art performance on NIH-AAPM dataset, similar to existing approaches. More importantly, we show that using such a sinogram extrapolation module significantly improves the generalization capability of the model on unseen datasets (e.g., COVID-19 and LIDC datasets) when compared to existing approaches.
		
		\keywords{Limited-Angle CT Reconstruction \and Sinogram Extrapolation \and Model Generalizability.}
	\end{abstract}
	
	\section{Introduction and Motivation}
	
	In healthcare, Computed Tomography(CT) based on X-ray projections is an indispensable imaging modality for clinical diagnosis. Limited-angle (LA) CT is a common type of acquisition in many scenarios, such as to reduce radiation dose in low-dose CT or forced to take projections in a restricted range of angles in C-arm CT~\cite{schafer2011mobile} and dental CT.
	However, the deficiency of projection angles brings significant challenge to image reconstruction and may lead to severe artifacts in the reconstructed images. 
	
	Many CT image reconstruction algorithms have been proposed in the literature to improve image quality, which can be categorized as model-based and deep-learning-based methods. For example, Filtered Back Projection (FBP)~\cite{wang2019machine}, as a representative analytical method, is widely used for reconstructing a high-quality image efficiently. However, FBP prefers acquisition with full-ranged views which makes using it for LACT sub-optimal. The (some times extreme) reduction on the range of projection angles decreases the effectiveness of the commercial CT reconstruction algorithms. To overcome such challenge, iterative regularization-based algorithms \cite{geman1995nonlinear,mahmood2018adaptive,rantala2006wavelet,sidky2008image,wang2017reweighted,zeng2015spectral} are proposed to leverage prior-knowledge on the image to be reconstructed and achieve better reconstruction performance for LACT. Notice that those iterative algorithms are often computationally expensive and require careful case-by-case hyperparameter tuning.
	
	Currently, deep learning(DL) techniques have been widely adopted in CT and demonstrate promising reconstruction performance~\cite{chen2017low,jin2017deep,yang2018low,zhang2020review,zhou2021review,zhou2019handbook}. By further combining the iterative algorithms with DL, a series of iterative frameworks with the accordingly designed neural-network-based modules are proposed~\cite{adler2018learned,cheng2020learned,ding2020low,gupta2018cnn,mardani2018neural,solomon2019deep,zhang2019jsr}. ADMMNet~\cite{sun2016deep} introduces a neural-network-based module in reconstruction problem and achieves remarkable performance. Furthermore, DuDoNet~\cite{lin2019dudonet,lyu2020encoding}, ADMM-CSNet~\cite{yang2018admm} and LEARN++~\cite{zhang2020learn++} improve reconstruction results with an enhancement module in the projection domain, which inspires us to fuse dual-domain learning in our model design.
	
	Although deep-learning-based algorithms have achieved state-of-the-art performance, they are also known to easily over-fit on training data, which is not expected in practice. MetaInvNet~\cite{zhang2020metainv} is then proposed to improve the reconstruction performance with sparse-view projections, demonstrating good model generalizability. They attempt to find better initialization for an iterative HQS-CG~\cite{geman1995nonlinear} model with a U-Net~\cite{ronneberger2015u} and achieve better generalization performance in such scenarios. But they still focus on the case with a large range of acquired projections, which limits the application of their model in practice. How to obtain a highly generalizable model when learning from practical data is still difficult.

	To retain model generalizability in LACT reconstruction, we propose a model, called ExtraPolationNetwork (EPNet), for recovering high-quality CT images. In this model, we utilize dual-domain learning to emphasize data consistency between image domain and projection domain, and introduce an extrapolation module. The proposed extrapolation module helps complement missed information in the projection domain and provides extra details for reconstruction. Extensive experimental results show that the model achieves state-of-the-art performance on the NIH-AAPM dataset~\cite{mccollough2016tu}. Furthermore, we also achieve better generalization performance on additional datasets, COVID-19 and LIDC~\cite{clark2013cancer}. This empirically verifies the effectiveness of the proposed extrapolation module. We make our implementation available at https://github.com/mars11121/EPNet.

	\section{Problem Formulation}
	CT reconstruction aims to reconstruct clean image $u$ from the projection data $Y$ with unknown noise $n$, whose mathematical formulation is:
	\begin{equation*}
	Y=Au+n,
	\end{equation*}
	where $A$ is the Radon transform. 
	For LACT, the projection data $Y$ is incomplete as a result of the decrease of angle range (the view angle $\alpha \in [0, \alpha_{max}]$ with $\alpha_{max} < 180\degree$). The reduced sinogram information limits the performance of current reconstruction methods. Therefore, the estimation of a more complete sinogram $\widetilde{Y}$ is necessary to enhance model reconstruction performance. To yield such an accurate estimation, the consistency between incomplete projection $Y$ and complete projection $\widetilde{Y}$ is crucial. We assume $Y$ is obtained by some operations (e.g. downsampling operation) from $\widetilde{Y}$. Besides, $\widetilde{Y}$ and clean image $u$ should also be consistent under the corresponding transformation matrix $\widetilde{A}$. Consequently, we propose the following constraints:
	\begin{equation}\label{data-consistency}
	\begin{split}
	\widetilde{A}u = \widetilde{Y}, 
	\quad P\widetilde{Y} = Y,
	\end{split} 
	\end{equation}
	where $P$ is the downsampling matrix. 
	
	In this way, the final model becomes the following optimization problem:
	\begin{equation}\label{problem}
	\begin{split}
	\min_{u} \quad& \frac{1}{2}\| Au-Y \|^{2} + R(u)\\
	s.t. \quad& \widetilde{A}u = \widetilde{Y}, 
	\quad P\widetilde{Y} = Y,
	\end{split} 
	\end{equation}
	where $R(\cdot)$ is a regularization term incorporating image priors.
	
	\section{Proposed Method}
\begin{figure}[t]
		\begin{center}
			\includegraphics[height=5.3cm]{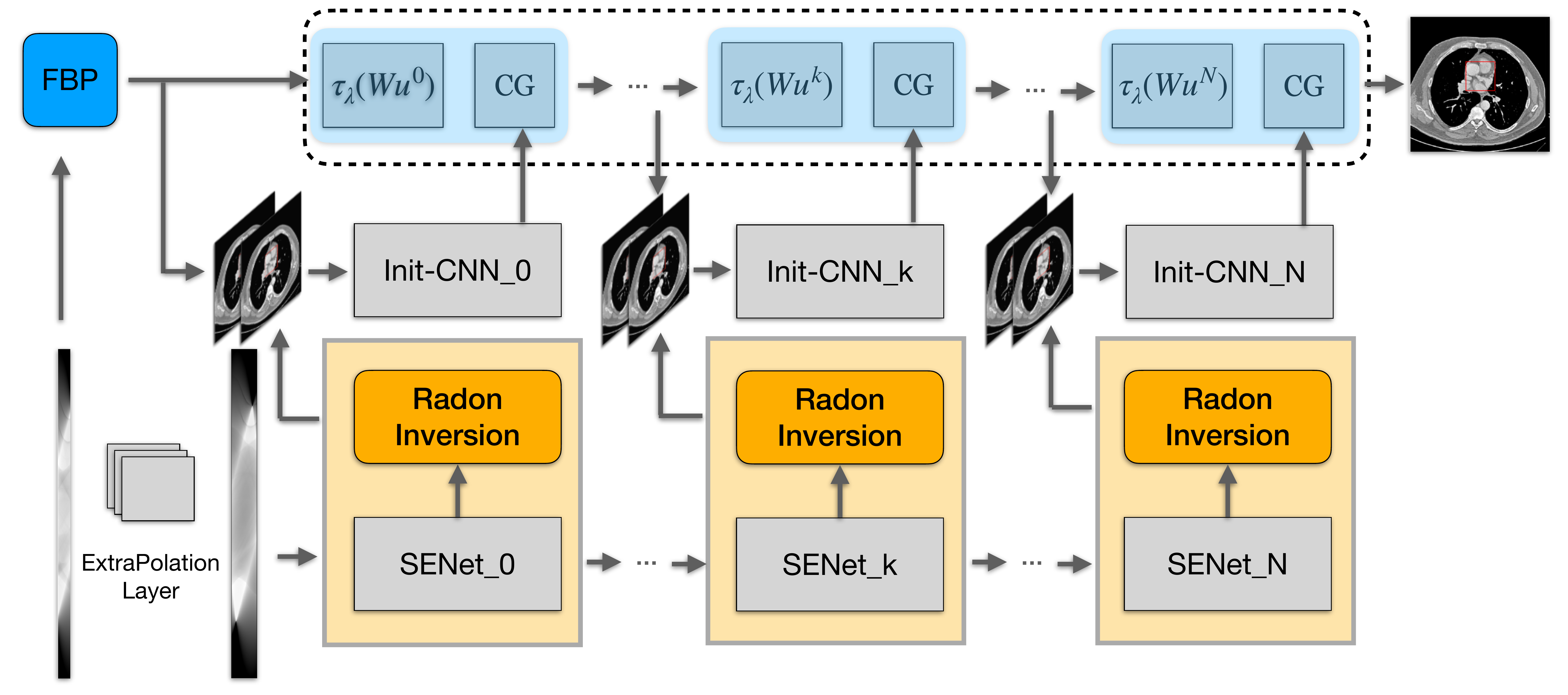}
			\caption{The framework of our proposed EPNet. It reconstructs CT image via two parallel pipelines in the image domain and sinogram domain. In the sinogram domain, we propose an extrapolation layer before SENet to take extra prior information of image details. The reconstructed images of sinogram domain and image domain are then concatenated followed by an "Init-CNN" module to provide a more accurate initialization estimation for the Conjugate-Gradient algorithm.}
			\label{framework}
		\end{center}
	\end{figure}

	In this section, we introduce full details on the proposed method, which is depicted in Fig.~\ref{framework}. 
	Our model is built by unrolling the HQS-CG~\cite{geman1995nonlinear} algorithm with $N$ iterations. The HQS-CG algorithm is briefly introduced in Section~\ref{sec.cg}.
	Specifically, we utilize the Init-CNN module~\cite{zhang2020metainv} to search for a better initialization for Conjugate Gradient (CG) algorithm in each iteration. The input of the module is composed of reconstructed images from the image domain and projection domain. In the image domain, we retrain the basic HQS-CG model and use the CG module for reconstruction. In the projection domain, we first use our proposed Extrapolation Layer (EPL) to estimate extra sinograms. Then, we use Sinogram Enhancement Network (SENet) to inpaint the extrapolated sinograms and reconstruct them with Radon Inversion Layer (RIL)~\cite{lin2019dudonet}, which is capable of backpropagating gradients to the previous layer. Section~\ref{main-method} introduces the details of the involved modules.
	\vspace{0.2cm}
	
	\subsection{HQS-CG Algorithm} \label{sec.cg}
	Traditionally, there exist many effective algorithms to solve objective~(\ref{problem}). One such algorithm is Half Quadratic Splitting(HQS)~\cite{geman1995nonlinear}, which solves the following:
	\begin{equation}\label{unconstrained-hqs-modify}
	\min_{u,z} \quad \frac{1}{2}\| Au-Y \|^{2} + \lambda\|z\|_{1} + \frac{1}{2}\sum_{i=1}^{M}{\gamma}_{i}\|{W}_{i}u - z_{i}\|^{2}+{\beta}_{1}\|P\widetilde{Y}-Y\|^{2} + {\beta}_{2}\|\widetilde{A}u - \widetilde{Y}\|^{2},
	\end{equation}
	where $W=({W}_{1},{W}_{2},\ldots,{W}_{M})$ is a $M-$channel operator, $z=({z}_{1},{z}_{2}, \ldots,{z}_{M})$, $\lambda >0 $, $\beta_{1}>0$, $\beta_{2}>0$, and $\gamma=({\gamma}_{1},{\gamma}_{2},\ldots,{\gamma}_{M})$ with $\{{\gamma}_{i}\}_{i=1}^{M} > 0$. The operator $W$ is chosen as the highpass components of the piecewise linear tight wavelet frame transform. With alternating optimization among $\widetilde{Y}$, $u$, and $z$, the final closed-form solution could be derived as follows:
	\begin{align}
	{\widetilde{Y}}^{k+1}&={\left( {\beta}_{1}{P}^{T}P+{\beta}_{2} \right)}^{-1}\left[  {\beta}_{1}{P}^{T}Y+{\beta}_{2}\widetilde{A}u^{k} \right],\notag\\
	{u}^{k+1}&={\left({A}^{T}A+\sum_{i=1}^{M}{\gamma}_{i}{{W}_{i}}^{T}{W}_{i}+2\beta_{2}\widetilde{A}^{T}\widetilde{A} \right)}^{-1} \left[ {A}^{T}Y+\sum_{i=1}^{M}{\gamma}_{i}{{W}_{i}}^{T}{{z}_{i}}^{k}+2\beta_{2}\widetilde{A}^{T}\widetilde{Y}^{k+1} \right],\notag\\
	{z}_{i}^{k+1}&=\tau_{\lambda/\gamma_{i}}(W_{i}{u}^{k+1}), i=1,...,M,\notag
	\end{align}
	where ${\tau}_{\lambda}(x) = \sign x\max\{\|x\|-\lambda, 0\}$ is the soft-thresholding operator.
	
	\subsection{Dual-domain Reconstruction Pipelines}\label{main-method}

	\noindent \textbf{Init-CNN and RIL.} We realize the Init-CNN module with a heavy U-Net architecture with skip connection, which stabilizes the training. Besides, the heavy U-Net shares parameters across different steps, which is proved more powerful for the final reconstruction. The Radon Inversion Layer (RIL) is first introduced in DuDoNet~\cite{lin2019dudonet}, which builds dual-domain learning for reconstruction. We here use the module to obtain the reconstructed image from the projection domain.
	
	\noindent \textbf{EPL.} As introduced, the reduction of angle range is the main bottleneck in the limited-angle scenario. Besides, usual interpolation techniques are not suitable in this case. But few researchers consider extrapolating sinograms with CNNs, which provides more details of images since sinograms contain both spatial and temporal (or view angle) information of the corresponding images. Compared with the image domain difference, sinograms from different data distributions also have similarities in the temporal dimension. To utilize such an advantage, we propose a module called ``Extrapolation Layer (EPL)" to extrapolate sinograms before SENet.
	
\begin{figure}[t]
		\begin{minipage}[b]{.5\linewidth}
			\begin{center}
				\includegraphics[height=3cm]{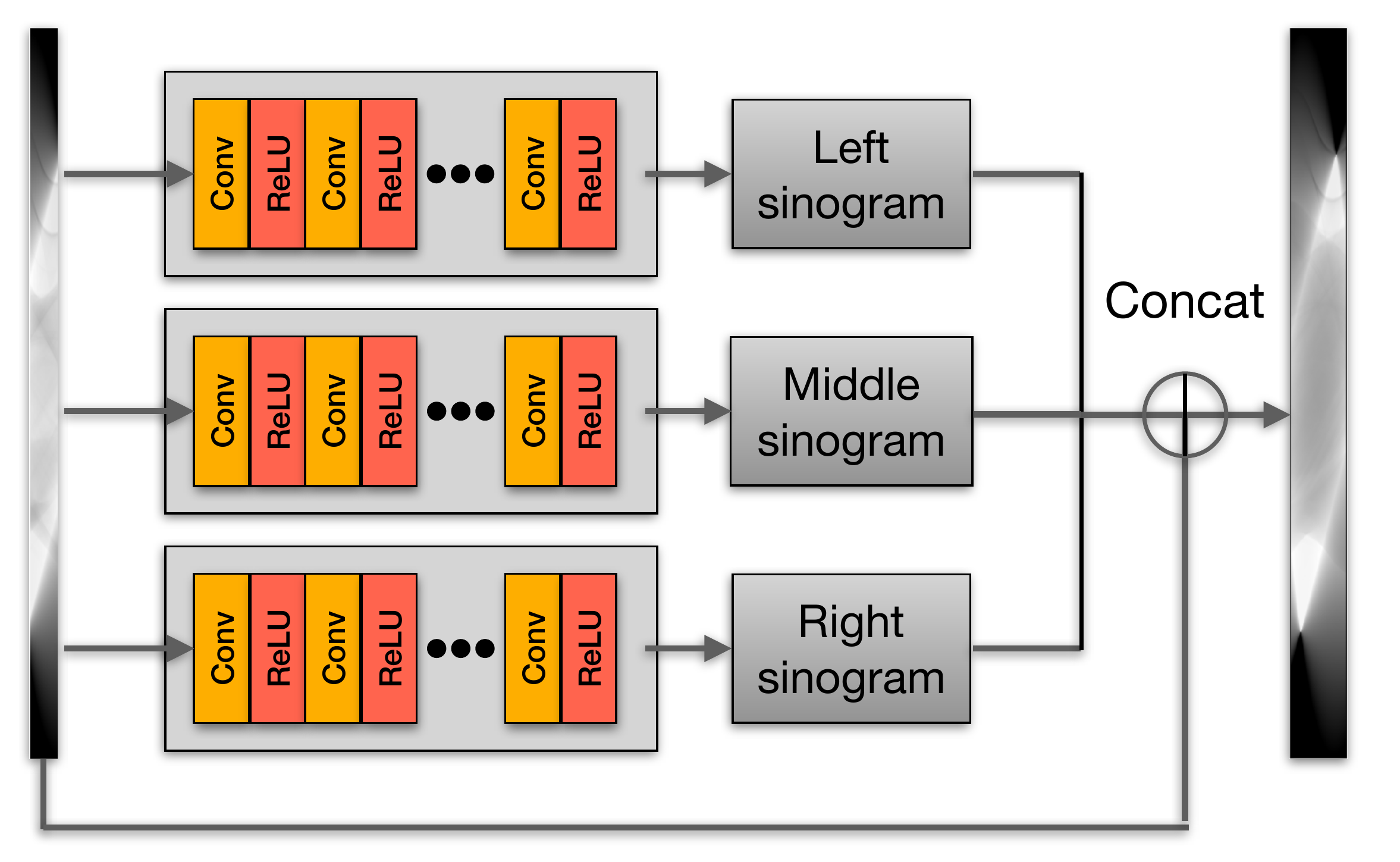}
				\caption{The architecture of EPL.}
				\label{epl}
			\end{center}
		\end{minipage}
		\begin{minipage}[b]{.5\linewidth}
			\begin{center}
				\includegraphics[height=3cm]{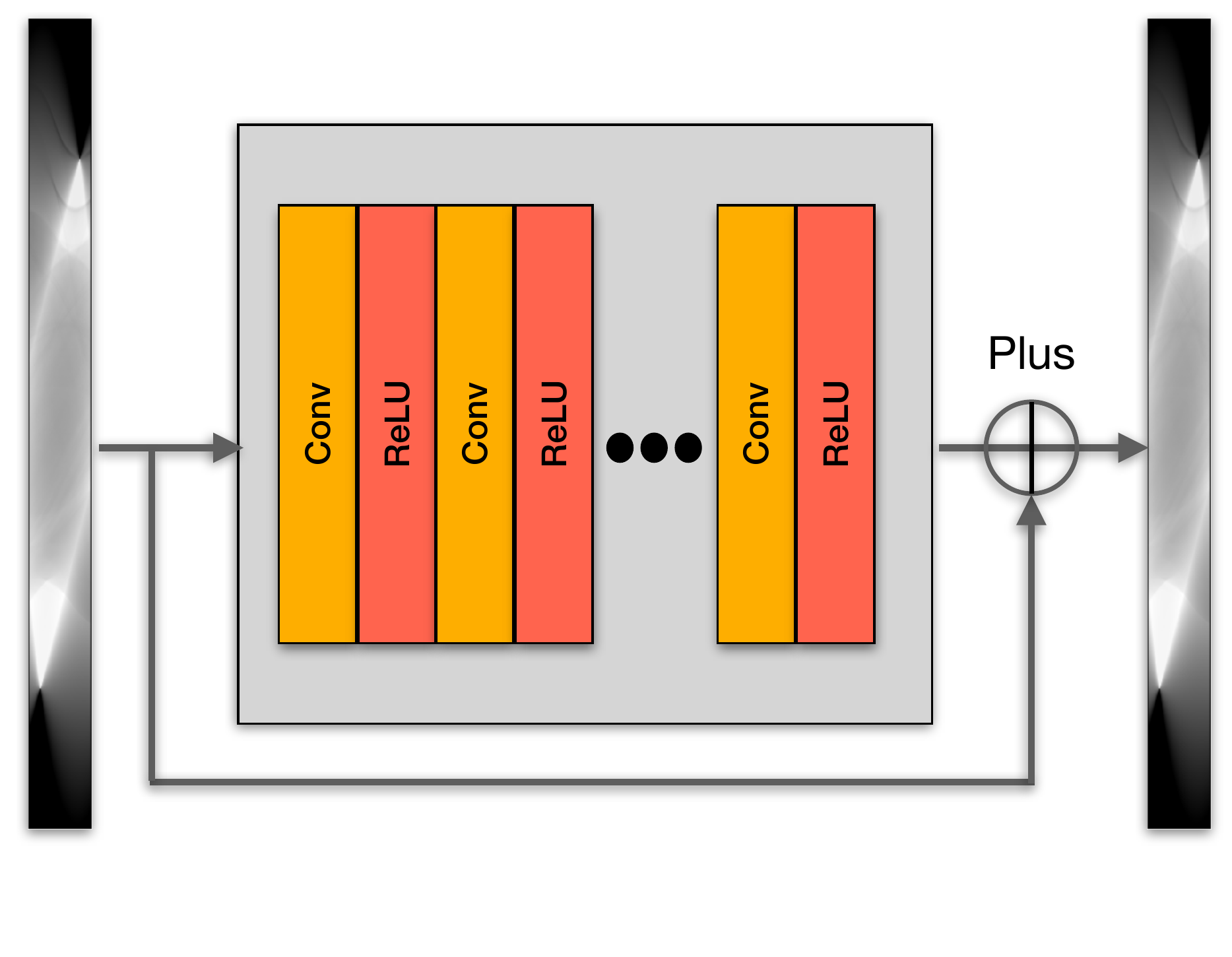}
				\caption{The architecture of SENet.}
				\label{SENet}
			\end{center}
		\end{minipage}
	\end{figure}

	As shown in Fig. \ref{epl}, the EPL module is composed of three parallel convolutional neural networks, where the left and the right networks are used to predict neighboring sinograms of the corresponding sides and the middle one is used to denoise the input. The outputs of the three networks are then concatenated, followed with the proposed  supervision defined as follows:
	\begin{equation}
	{\cal L}_{EPL} = (1+mask) \times \| {Y}_{out} - {Y}_{gt} \|_{1} + \| RIL({Y}_{out})-RIL({Y}_{gt}) \|_{1},
	\end{equation}
	where ${Y}_{out}$ is the predicted sinogram, ${Y}_{gt}$ is the corresponding ground-truth, and $mask$ is a binary matrix to emphasize the bilateral prediction. Here, we utilize RIL to realize a dual-domain consistency for the prediction, which makes the module estimation more accurate when embedded into the whole model.
	
	\noindent \textbf{SENet.} With extrapolated sinograms, we then use SENet to firstly enhance the quality of sinograms, which is designed as a light CNN as in Fig. \ref{SENet}. At last, the enhanced sinograms are mapped to the image domain via RIL, which would help decrease the different optimization directions in our dual-domain learning. The objective for SENet is as follows:
	
	\begin{equation}
	{\cal L}_{SE} = \| {Y}_{se} - {Y}_{gt} \|_{1} + \| RIL({Y}_{se}) - {u}_{gt} \|_{1},
	\end{equation}
	where ${Y}_{se}$ is the enhanced sinogram, ${Y}_{gt}$ and ${u}_{gt}$ are the corresponding ground-truth sinogram and image, respectively. 
	
	\noindent \textbf{Loss function.} With the above modules, the full objective function of EPNet is defined by:
	\begin{equation}
	{\cal L} = \sum_{i=1}^{N}\| {u}_{i} - {u}_{gt} \|_{2} + \mu{\cal L}_{ssim}({u}_{N},{u}_{gt}) + {\cal L}_{EPL}+{\cal L}_{SE} ,
	\end{equation}
	where $N$ is the total iterations of unrolled back-bone HQS-CG model, $\{u\}_{i=1}^{N}$ is the reconstructed image of each iteration, and ${\cal L}_{ssim}$ is the SSIM loss.

	\section{Experimental Results}
	
	\begin{table}[t]
		\begin{center}
			\setlength{\abovecaptionskip}{0cm}
			\setlength{\belowcaptionskip}{-0.1cm}
			\caption{Quantitative results of models on different testing datasets. The best performances in each row are in \textbf{bold}.}
			\resizebox{\linewidth}{!}{ 
				\label{ablation}
				\begin{tabular}{| l | rr | rr | rr | rr | rr |rr|}
					\thickhline
					\multirow{2}*{{Ablation Study}}& \multicolumn{2}{c|}{{EPL30}} & \multicolumn{2}{c|}{{EPL30$_{re}$}} &\multicolumn{2}{c|}{{EPL60}}   &\multicolumn{2}{c|}{{EPL120}} & \multicolumn{2}{c|}{{ DuDoNet }} & \multicolumn{2}{c|}{{ DuDoEPL30 }} \\ 
					\cline{2-13}
					& PSNR & SSIM& PSNR &SSIM&PSNR &SSIM&PSNR& SSIM &PSNR& SSIM & PSNR&SSIM\\ 			\hline
					AAPM-test  & 27.63 &0.882 & \textbf{28.03}&\textbf{0.886}   &25.41  &0.867  &27.87&0.885 &23.71&0.716&22.63 &0.842  \\ 
					COVID-test & \textbf{19.59} &\textbf{0.713} & 18.11&0.649   &18.68  &0.674  &16.90&0.584 &5.59&0.252&6.94   &0.295   \\ 
					LIDC-test    & \textbf{19.80} &\textbf{0.726} & 18.81&0.670   &19.19  &0.690  &18.07&0.627 &5.45&0.246&6.93   &0.304  \\ \thickhline
				\end{tabular}
			}
		\end{center}
	\end{table}
	
	\subsection{Datasets and Experimental Settings}
	\textbf{Datasets.} We first train and test models on the ``2016 NIH-AAPM-Mayo Clinic Low Dose CT Grand Challenge" dataset \cite{mccollough2016tu}. Specifically, we choose 1,746 slices of five patients for training and 1,716 slices of another five patients for testing. To further show our models' generalization capability, we test our models on 1,958 slices of four patients chosen from the COVID-19 dataset, and 1,635 slices of six patients from the LIDC dataset~\cite{clark2013cancer}. The two datasets are also composed of chest CT images but from different scenarios and machines, which constitutes good choices for testing the generalization capability. For the generalizability experiments, we have added additional HU value shift to the latter two datasets, which brings about similar distribution deformation. All the experiments are conducted with Fan-Beam Geometry and the number of detector elements is set to 800. Besides, we add mixed noise, composed of 5\% Gaussian noise and Poisson noise with an intensity of $5{e}^{6}$, to all simulated sinograms~\cite{zhang2020metainv}.
	
	\noindent \textbf{Implementations and Training Settings.} All the compared models are trained and tested with the corresponding angle number (15, 30, 60, 90) except MetaInvNet\_ori, which is trained with 180 angle number as Zhang et al. \cite{zhang2020metainv} do. Our models are implemented using the PyTorch framework. We use the Adam optimizer~\cite{kingma2014adam} with $({\beta}_{1}, {\beta}_{2})$ = (0.9, 0.999) to train these models. The learning rate starts from 0.0001. Models are all trained on a Nvidia 3090 GPU card for 10 epochs with a batch size of 1.
	
	\noindent \textbf{Evaluation Metric.} Quantitative results are measured by the multi-scale structural similarity index (SSIM) (with level=5, Gaussian kernel size=11, and standard deviation=1.5) \cite{wang2003multiscale} and peak signal-to-noise ratio (PSNR) \cite{wang2004image}.
	
	\subsection{Ablation Study}
	To investigate the effectiveness of different modules and used hyperparameters for models, we firstly conduct an ablation study with the following configurations, where the number of the input sinogram angle is fixed to $\alpha_{max}=60$: \\
	\noindent a) EPL30: our model with pretrained EPL fixed and extrapolate 30 angles, 
	\\ \noindent b) {EPL30}$_{re}$: our model with pretrained EPL not fixed and extrapolate 30 angles, 
	\\ \noindent c) EPL60: our model with pretrained EPL fixed and extrapolate 60 angles,
	\\ \noindent d) EPL120: our model with pretrained EPL fixed and extrapolate 120 angles,
	\\ \noindent e) DuDoEPL30: DuDoNet with our proposed EPL and extrapolate 30 angles.
	
	\begin{table}[t]
		\begin{center}
			\setlength{\abovecaptionskip}{0cm}
			\setlength{\belowcaptionskip}{0.15cm}
			\caption{Quantitative results on AAPM-test, COVID-test and LIDC-test sets. We also test the computation time when sinogram number is fixed as 60.}
			\resizebox{0.9\linewidth}{!}{
				\begin{tabular}{| l | rr | rr | rr | rr | r|}
					\thickhline
					\multirow{2}{*}{{AAPM-test}}& \multicolumn{2}{c|}{{ $\alpha_{max}$ = 15}}  &\multicolumn{2}{c|}{{$\alpha_{max}$ = 30}}   &\multicolumn{2}{c|}{{$\alpha_{max}$ = 60}} & \multicolumn{2}{c|}{{$\alpha_{max}$ = 90}} & \multirow{2}{*}{{Time(s)}} \\ 
					\cline{2-9}
					& PSNR & SSIM& PSNR &SSIM&PSNR &SSIM&PSNR& SSIM & \\ \hline
					FBP            &7.12   &0.399 &8.61   &0.495    &12.63    &0.643    &14.98     &0.726 & 0.0123   \\ 
					HQS-CG     &18.66&0.621& 19.95  &0.675  &21.65    &0.744   & 23.96      &0.800 & 15.7620\\ 
					DuDoNet    &20.04  &0.633&21.11 &0.661     &23.71    &0.716    &24.95    &0.753  &  0.1321  \\ 
					FBPCovNet  &20.38  &0.802&22.04 &0.831     &23.69    &0.855    &28.09    &0.897  &  0.1445  \\  
					MetaInvNet\_ori &17.60 &0.800&18.99  &0.822    &21.47    &0.856    &24.04     &0.884 &   1.6528   \\
					MetaInvNet &\textbf{21.99}&\textbf{0.819} & \textbf{24.03}     &\textbf{0.845}  &\textbf{28.21}      &\textbf{0.887}   &\textbf{30.05}   &\textbf{0.902}    &  1.5885  \\ \hline
					EPNet &\textbf{21.92} &\textbf{0.820}&\textbf{23.65}  &\textbf{0.842}   &\textbf{27.63}     &\textbf{0.882}    &\textbf{30.40}     &\textbf{0.906} & 1.8859   \\ 	\hline\hline
					COVID-test& PSNR & SSIM& PSNR &SSIM&PSNR &SSIM&PSNR& SSIM & Time(s) \\ \hline
					FBP            &8.24   &0.4073& 9.31   &0.4516    & 11.19    &0.5177     &12.98    &0.5688  &   0.0101    \\ 
					HQS-CG     &\textbf{17.84} &\textbf{0.611}& \textbf{19.29} & \textbf{0.679}   &\textbf{21.18}      &\textbf{0.782 }    &\textbf{23.41}    &\textbf{0.824} & 16.6641  \\ 
					DuDoNet    & 4.15  &0.214&4.29    &0.231    &5.59        &0.252     &6.26      &0.268 &  0.1336  \\ 
					FBPCovNet  & 3.96  &0.273&3.88    &0.279    &4.91       &0.296      &6.60      &0.328 &   0.1418  \\  
					MetaInvNet\_ori &\textbf{16.03} &\textbf{0.565}&16.80  &0.584   & 18.45     &0.644     &20.86     &0.719  &   1.8310   \\
					MetaInvNet &13.05 &0.339&15.30     &0.298    &15.31       &0.531   &18.37    &0.591    &  1.8199  \\ \hline
					EPNet &15.56 &0.508&\textbf{17.31}  &\textbf{0.616}   &\textbf{19.59}      &\textbf{0.713}     &\textbf{21.57}     &\textbf{0.745}  &  1.9004  \\ 	\hline\hline
					LIDC-test& PSNR & SSIM& PSNR &SSIM&PSNR &SSIM&PSNR& SSIM & Time(s) \\ \hline
					FBP            &8.57    &0.439&9.57   &0.492    &11.52     &0.551      &13.48   &0.595   &   0.0116    \\ 
					HQS-CG     &\textbf{18.71}  &\textbf{0.602}&\textbf{20.30} &\textbf{0.673}    &\textbf{22.27}     & \textbf{0.787}     & \textbf{24.46}   &\textbf{0.837}  &  16.5000   \\ 
					DuDoNet    &4.12    &0.205&4.19  &0.223     &5.45      &0.246       &6.24     &0.266    &  0.1316 \\ 
					FBPCovNet  &3.89    &0.276&3.80  &0.284     &4.89      &0.303       &6.74     &0.343  & 0.1407   \\  
					MetaInvNet\_ori &\textbf{16.59} &\textbf{0.561}&17.40  &0.581    &18.74     &0.634      & 20.75  &0.698  &   1.7403   \\
					MetaInvNet &13.57 &0.367&15.66      &0.517   &15.45        &0.547 &18.33    &0.583    &  1.7616  \\ 	\hline	
					EPNet &16.07 &0.564&\textbf{18.02} &\textbf{0.630}    &\textbf{19.80}      &\textbf{0.726}      &\textbf{22.18}   &\textbf{0.751} &  1.9180  \\ 	\thickhline
				\end{tabular}
			}
		\end{center}
	\end{table}
	
	\noindent The quantitative comparison is shown in Table \ref{ablation}. When comparing (a) and (b), retraining parameters of the pretrained EPL module reduces the generalizability of our model, so we \underline{fix} the EPL module parameters in later experiments. Besides, we investigate the most suitable extrapolated angle number for EPL. Comparing models (a) (c) (d), when increasing the number of extrapolated angles from 30 to 120, the reconstruction performance on AAPM-test is not affected, but the generalization performance gradually reduces. Therefore, we fix the number of extrapolated angles as 30 in all later experiments. Besides, we also insert the module in DuDoNet~\cite{lin2019dudonet}, the reconstruction performance drops a lot, but the module also improves generalization result by about 1.5 dB.
	
	\subsection{Quantitative and Qualitative Results Comparison}
	\noindent\textbf{Quantitative Results Comparison.} Then, we quantitatively compare our models with model-based and data-driven models. Results on the AAPM-test set show that the performance of our models and retrained MetaInvNet~\cite{zhang2020metainv} are the best. Besides, the original training setting of MetaInvNet has achieved a better generalization performance on COVID-test and LIDC-test sets, but they need more projections to train the model and our models have also achieved better generalizability results than it except when $\alpha_{max}$ = 15, which is due to the extremely limited sinogram information fed into extrapolation layer. On the other hand, HQS-CG has kept their performance across different data distributions, however the prior knowledge modeling limits their reconstruction performance on AAPM-test set, and the tuning and computation time is too expensive.
	
	\begin{figure*}[!t]
		\begin{minipage}[t]{0.1375\textwidth}
			\centering
			\includegraphics[width=\textwidth]{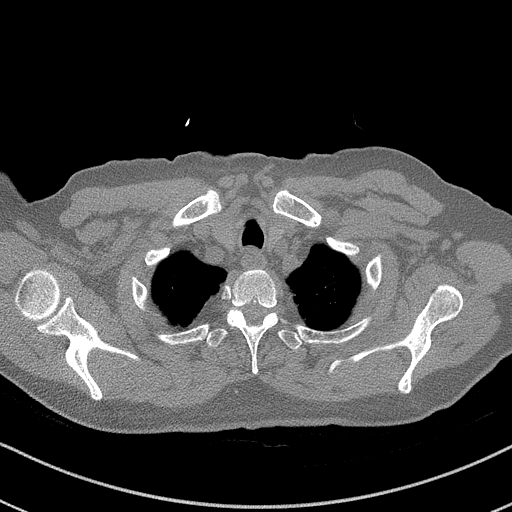}
			\includegraphics[width=\textwidth]{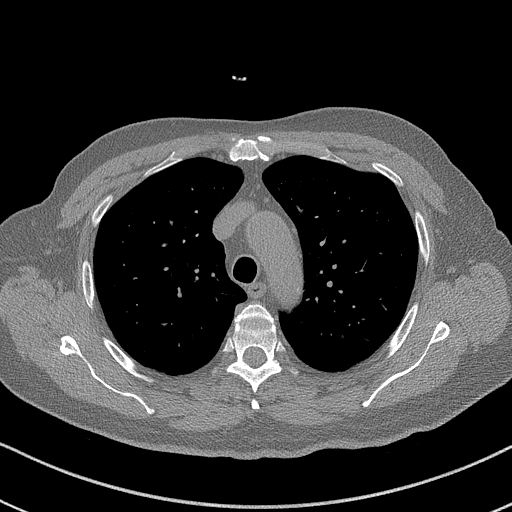}
			\includegraphics[width=\textwidth]{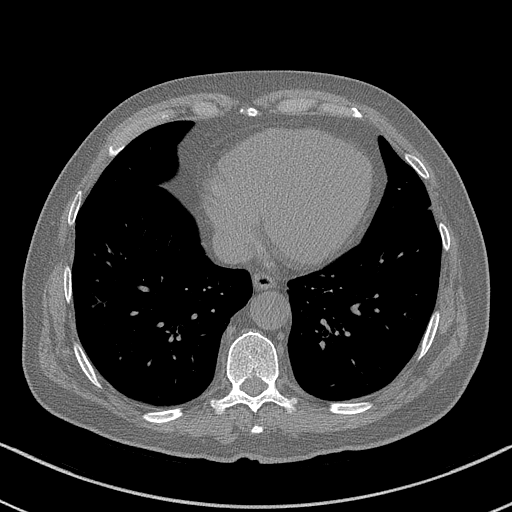}
			\includegraphics[width=\textwidth]{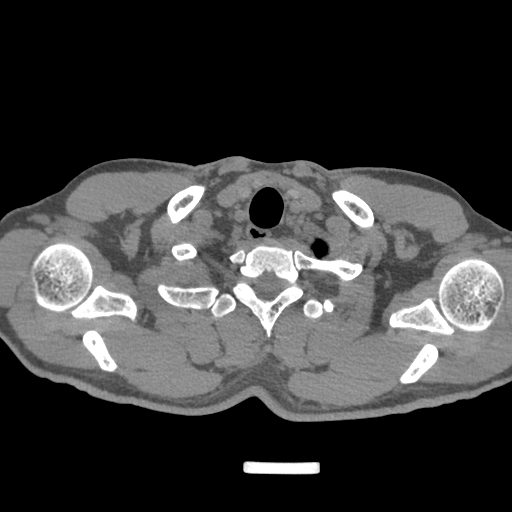}
			GT
		\end{minipage}
		\begin{minipage}[t]{0.1375\textwidth}
			\centering
			\includegraphics[width=\textwidth]{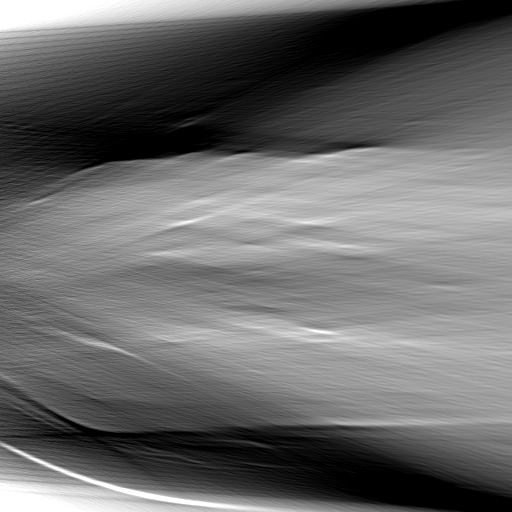}
			\includegraphics[width=\textwidth]{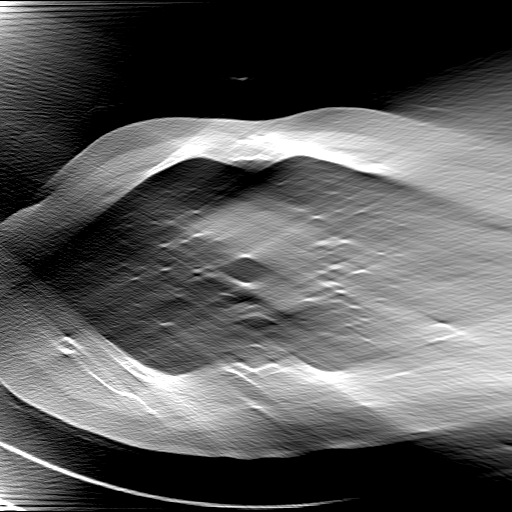}
			\includegraphics[width=\textwidth]{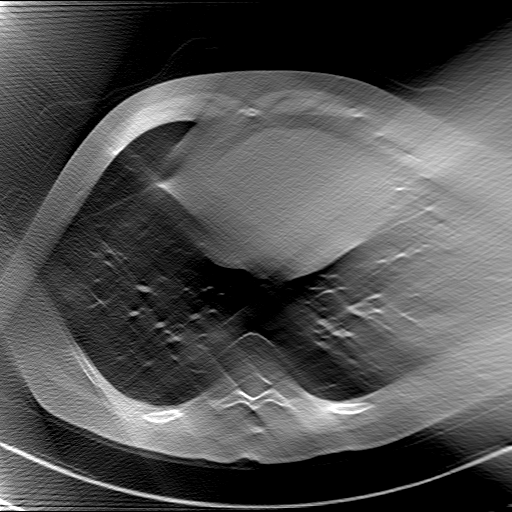}
			\includegraphics[width=\textwidth]{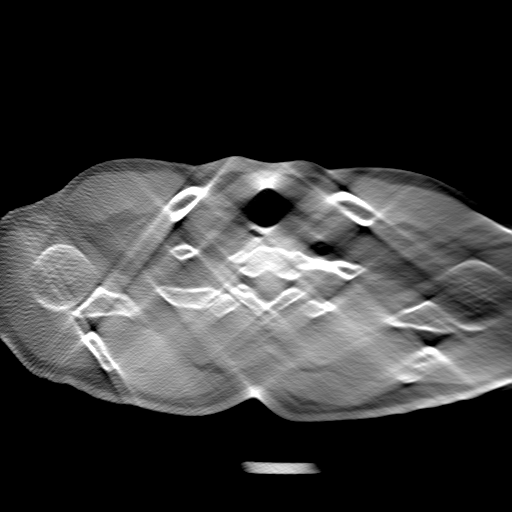}
			FBP
		\end{minipage}
		\begin{minipage}[t]{0.1375\textwidth}
			\centering
			\includegraphics[width=\textwidth]{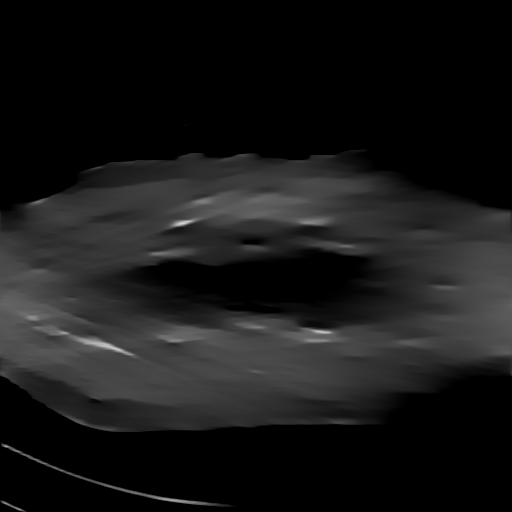}
			\includegraphics[width=\textwidth]{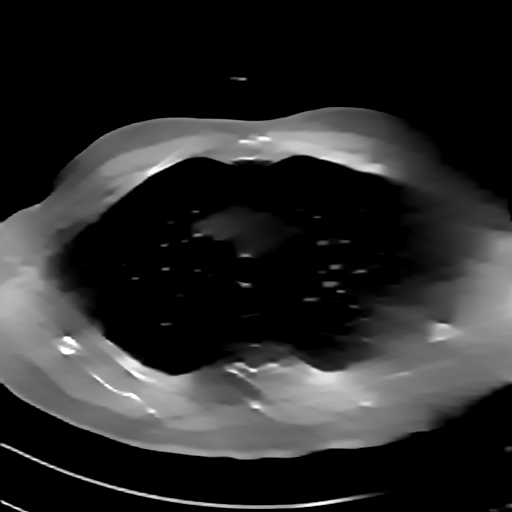}
			\includegraphics[width=\textwidth]{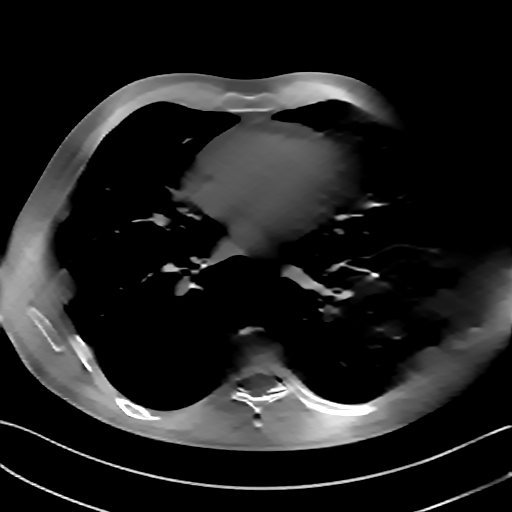}
			\includegraphics[width=\textwidth]{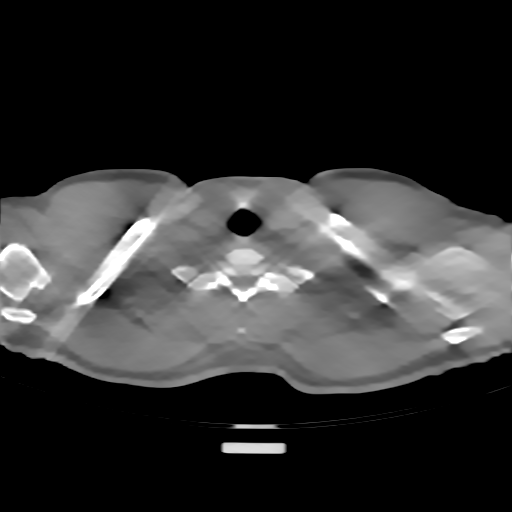}
			HQS-CG
		\end{minipage}
		\begin{minipage}[t]{0.1375\textwidth}
			\centering
			\includegraphics[width=\textwidth]{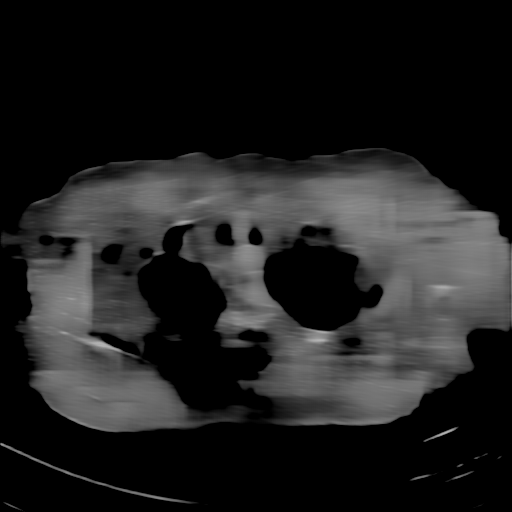}
			\includegraphics[width=\textwidth]{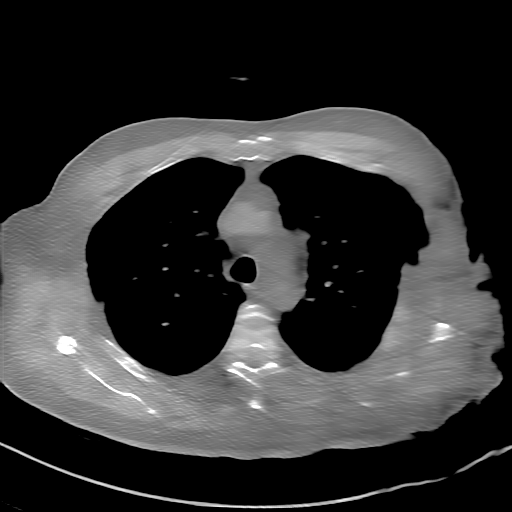}
			\includegraphics[width=\textwidth]{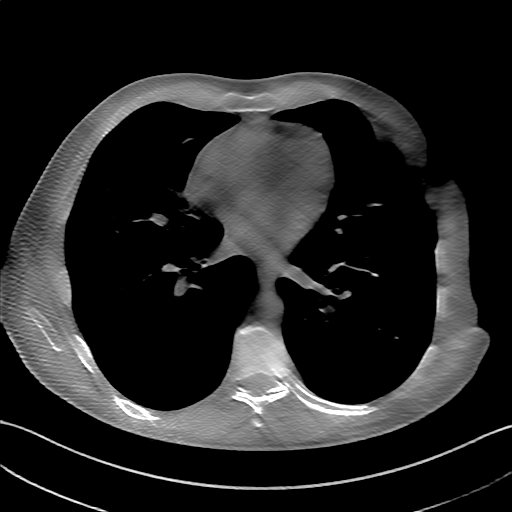}
			\includegraphics[width=\textwidth]{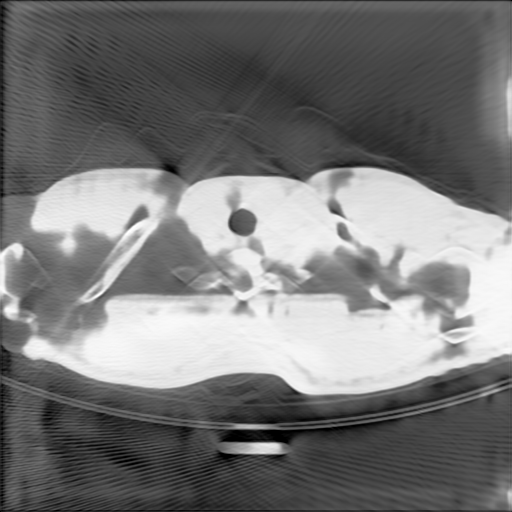}
			DuDoNet
		\end{minipage}
		\begin{minipage}[t]{0.1375\textwidth}
			\centering
			\includegraphics[width=\textwidth]{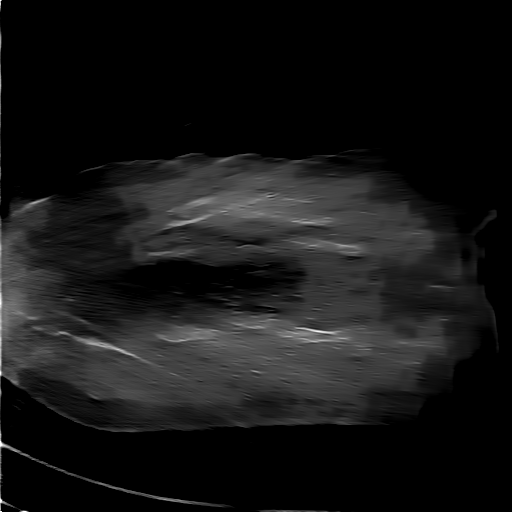}
			\includegraphics[width=\textwidth]{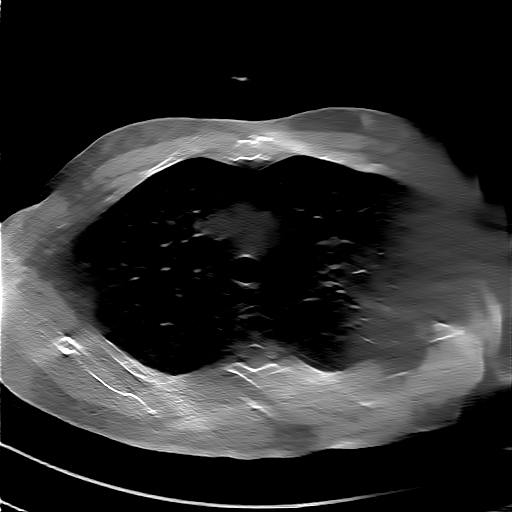}
			\includegraphics[width=\textwidth]{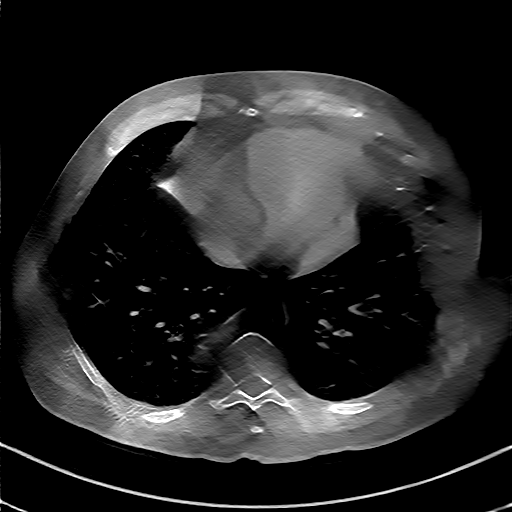}
			\includegraphics[width=\textwidth]{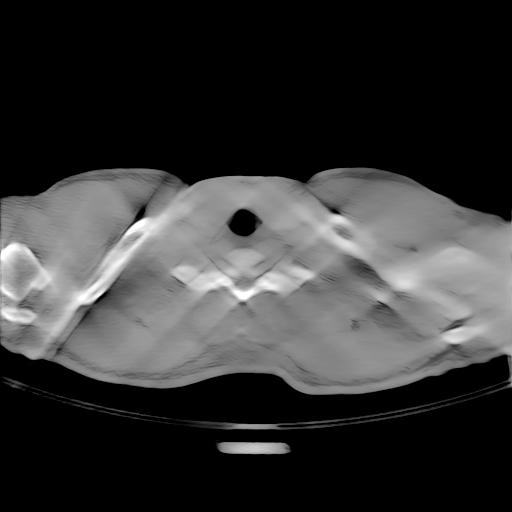}
			Meta
		\end{minipage}
		\begin{minipage}[t]{0.1375\textwidth}
			\centering
			\includegraphics[width=\textwidth]{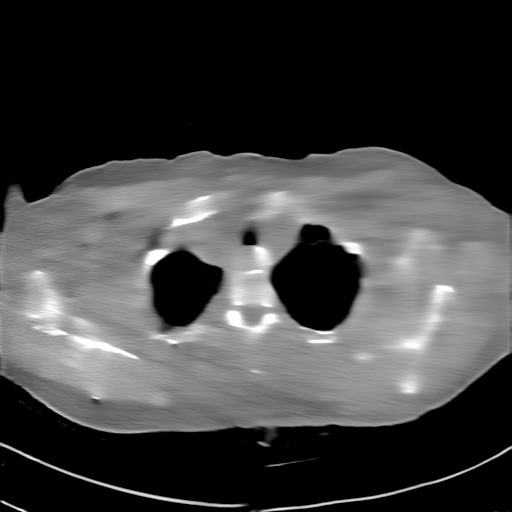}
			\includegraphics[width=\textwidth]{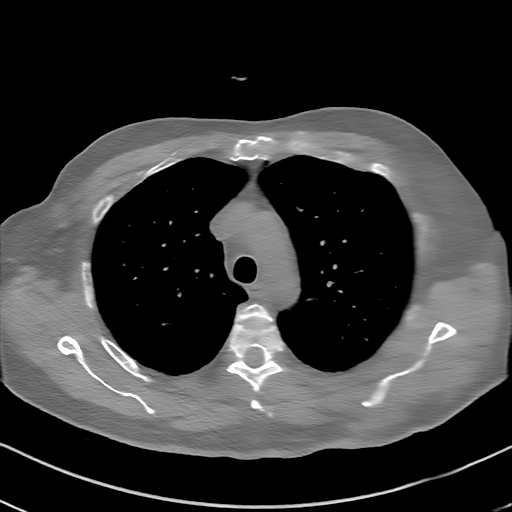}
			\includegraphics[width=\textwidth]{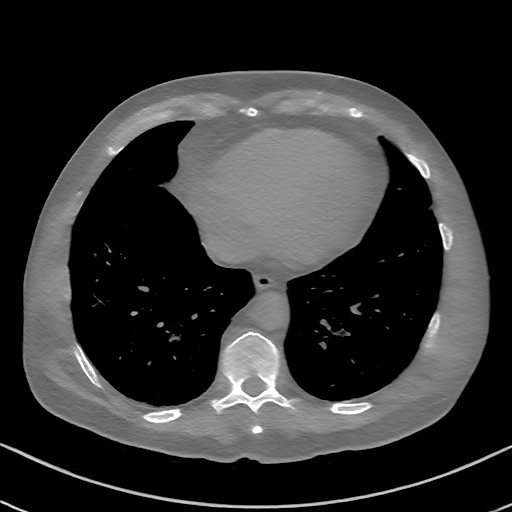}
			\includegraphics[width=\textwidth]{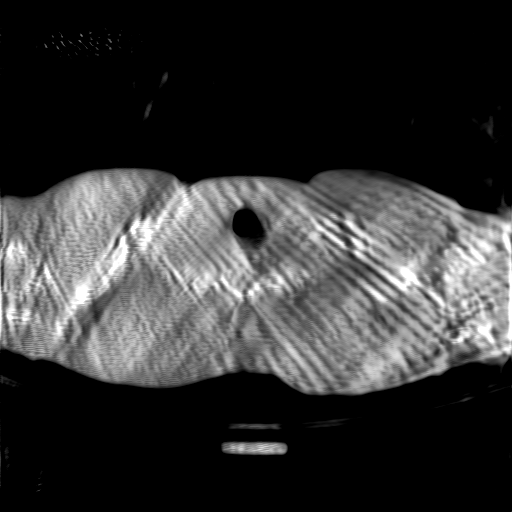}
			Meta\_re
		\end{minipage}
		\begin{minipage}[t]{0.1375\textwidth}
			\centering
			\includegraphics[width=\textwidth]{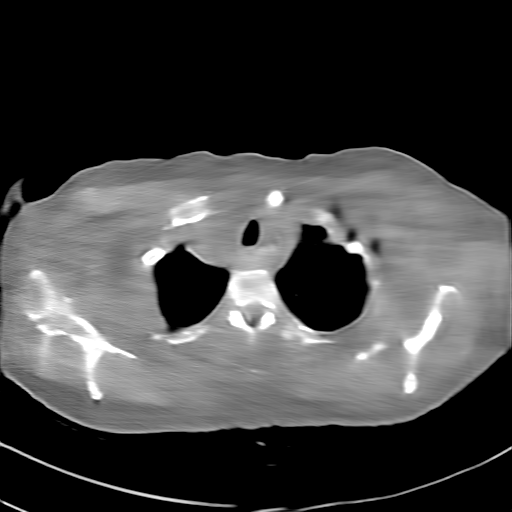}
			\includegraphics[width=\textwidth]{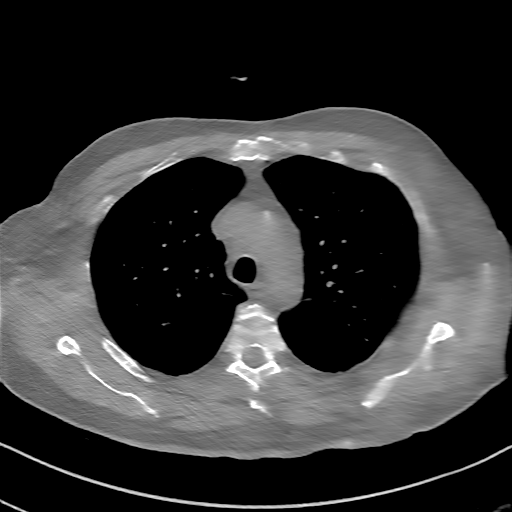}
			\includegraphics[width=\textwidth]{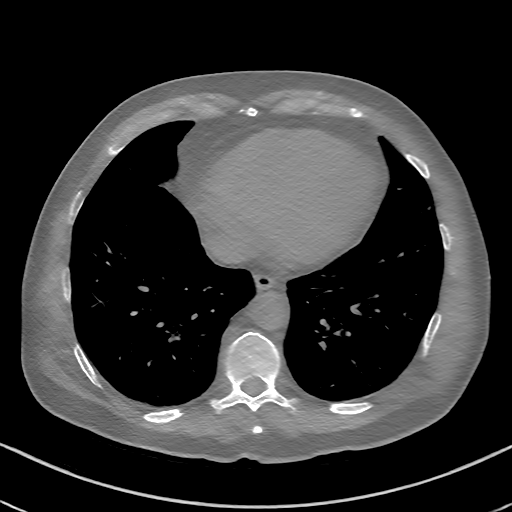}
			\includegraphics[width=\textwidth]{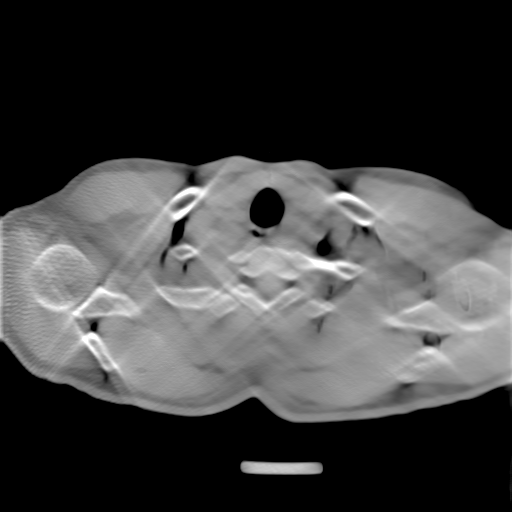}
			EPNet
		\end{minipage}
		\caption{The visualization of compared methods. The first three rows are results on AAPM-test set with 30, 60, and 90 input sinograms. The last row shows results on COVID-test set with 90 sinograms.}
		\label{vis-result}
	\end{figure*}
	
	\noindent\textbf{Qualitative Results Comparison.} We also visualize the reconstruction results of these methods on AAPM-test and COVID-test datasets. As in the first three rows of Fig. \ref{vis-result}, the reconstructed images from ours and retrained MetaInvNet show the best visualization quality on AAPM-test set across different angle numbers. Besides, our results show sharper details with the additional utilization of ${\cal L}_{SE}$ in the projection domain. When testing the reconstructed image on the COVID-test set, our result also gives sharper details but with more artifacts since the data distribution is very different. Although HQS-CG has achieved better quantitative results on the COVID-test dataset, the reconstructed image of their model in the fourth row is even smoother than FBP. 
	
	\section{Conclusion}
	We propose the novel EPNet for limited-angle CT image reconstruction and the model achieves exciting generalization performance. We utilize dual-domain learning for data consistency in two domains and propose an EPL module to estimate extra sinograms, which provide useful information for the final reconstruction. Quantitative and qualitative comparisons with competing methods verify the reconstruction performance and the generalizability of our model. The effectiveness encourages us to further explore designing a better architecture for EPL in the future.	
	
	\vspace{0.3cm}
	\noindent \textbf{Acknowledge.}
	This work was supported in part by the National Natural Science Foundation of China (NSFC) under Grant 11831002, in part by the Beijing Natural Science Foundation under Grant 180001, in part by the NSFC under Grant 12090022, and in part by the Beijing Academy of Artificial Intelligence (BAAI).
	
	%
	%
	%
	\bibliographystyle{splncs04}
	\bibliography{egbib}
\end{document}